\documentclass[11pt]{article}
\usepackage{latexsym}
\usepackage{epsfig}

\def\bg#1{\mbox{\boldmath$#1$}}

\newcommand{\anabla}{{\overrightarrow{\nabla}}\!\!\!\!\!\!{\overleftarrow{\nabla}}}

\newcommand{\del}{\partial}
\newcommand{\beq}{\begin{eqnarray}}
\newcommand{\eeq}{\end{eqnarray}}
\newcommand{\be}{\begin{eqnarray*}}
\newcommand{\ee}{\end{eqnarray*}}
\newcommand{\bk}{{\bf k}}
\newcommand{\bp}{{\bf p}}
\newcommand{\bq}{{\bf q}}
\newcommand{\br}{{\bf r}}

\newcommand{\ra}{\rightarrow}
\newcommand{\e}{\epsilon}

\newcommand{\nn}{\nonumber}
\newcommand{\ket}[1]{\mbox{$\mid\!#1\rangle$}}
\newcommand{\bra}[1]{\mbox{$\langle#1\!\mid$}}

\begin{document}

\centerline{\Large\bf {Proton-Proton Fusion in Effective Field Theory}}
\vskip 5mm
\centerline{Xinwei Kong}
\centerline{\it TRIUMF, 4004 Wesbrook Mall, Vancouver, B.C., Canada V6T 2A3.}
\vskip 5mm
\centerline{Finn Ravndal}
\centerline{\it Institute of Physics, University of Oslo, N-0316 Oslo, Norway.}

\bigskip
{\bf Abstract:} {\small The rate for the fusion process $p + p \ra d + e^+ + \nu_e$ is
calculated using non-relativistic effective field theory. Including the four-nucleon derivative
interaction, results are obtained in next-to-leading order in the momentum expansion. This
reproduces the effects of the effective range parameter. Coulomb interactions between the 
incoming protons are included non-perturbatively in a systematic way. The resulting fusion rate 
is independent of specific models and wavefunctions for the interacting nucleons. At this order in the
effective Lagrangian there is an unknown counterterm which limits the numerical accuracy of
the calculated rate given by the squared reduced matrix element $\Lambda^2(0) = 7.37$. Assuming the 
counterterm to have a natural magnitude, we estimate the accuracy of this result to be 6\% - 8\%. This
is consistent with previous nuclear physics calculations based on effective range theory and inclusion 
of axial two-body weak currents. The true magnitude of the counterterm can be determined from a precise 
measurement of the cross-section for low-energy neutrino scattering on deuterons.}

\section{Introduction}
One of the most important problems in modern physics is the nature and properties of neutrinos.
These were for a long time thought to be massless and stable, but experiments during the last decade have
consistently shown this to be incompatible with the observed neutrino oscillations\cite{Bahcall}.
Historically and even today the fusion processes in the Sun are among the few available and abundant 
sources of low-energy neutrinos available for experimental investigations. In order to study oscillations
in the detected fluxes, one needs to be sure of the production rates in the different nuclear reactions 
taking place in the Sun. 

The basic process is proton-proton fusion $p + p \ra d + e^+ + \nu_e$. It was explained more than 
sixty years ago by Bethe and Critchfield when nuclear physics was still at a very primitive
stage\cite{BetheC}. When the field had matured, it was reconsidered in the light of more modern 
developments by 
Salpeter who included effective range corrections\cite{ES}. Applications to the specific conditions
we have in the Sun were investigated by Bahcall and May\cite{BM}. This work was later extended by 
Kamionkowski and Bahcall who also included the effects of vacuum polarization in the Coulomb
interaction between the incoming protons\cite{KB}.
In spite of the enormous progress in nuclear physics during this time,
the methods and approximations made in these different calculations were essentially the 
same with a resulting accuracy in the fusion rate of a few percent. Including strong corrections due to
mesonic currents at smaller scales, the uncertainty in the rate is now around one percent\cite{Stoks}. 
This is very impressive for a strongly interacting process at low energies where ordinary perturbation 
theory cannot be used. 

In the light of the importance this fundamental process plays in connection with the solar
neutrino production and possible neutrino oscillations, it is natural to
reconsider the process from the point of view of modern quantum field theory instead of
the old potential models used previously. A first attempt in this direction was
made by Ivanov et al.\cite{Ivanov}. In their relativistic model they obtained a result 
which was significantly different from the standard result based upon conventional 
nuclear physics models. Subsequently it was pointed 
out by Bahcall and Kamionowski that their effective nuclear interaction was not 
consistent with what is known about proton-proton scattering at low energies where 
Coulomb effects are important\cite{BK_2}. In a more recent contribution this defect of their 
calculation was removed and better agreement with standard results have been obtained\cite{Ivanov_2}.

The approach of Ivanov et al. is based upon relativistic field theory and 
should in principle yield reliable results. But it is well known that it is very difficult to use 
consistently a relativistic formulation  for bound states like the deuteron. In addition, the 
fusion process considered here takes place at low energies and should therefore  instead be 
described within a non-relativistic framework. Then all the large-momentum degrees of freedom 
are integrated out
and one is left with an effective theory involving only the physically important field 
variables. The underlying, relativistic interactions are replaced by non-renormalizable 
local interactions with coupling constants which must be determined from experiments 
at low energies.  Along these lines the proton-proton fusion rate has been calculated
by Park, Kubodera, Min and Rho using chiral perturbation theory in the low-energy 
limit\cite{Park}. They obtain results in very good agreement with previous nuclear physics calculations. 
This is to be expected since they make use of phenomenological nucleon wavefunctions which fit low-energy 
scattering data very well. The drawback is that the results cannot be derived in an entirely analytical
way.

A more fundamental approach to nucleon-nucleon interactions at low energies has been formulated by Kaplan, 
Savage and Wise in terms of an effective theory for non-relativistic nucleons\cite{KSW_1}\cite{KSW_2}. 
It involves a few basic coupling constants which have been determined from nucleon 
scattering data at low energies. With no more free parameters to fit it can then be used to make predictions 
for a large number of other experimentally accessible quantities\cite{EFTbook}. The effects of pions can 
be included using the established counting rules and higher order corrections can be derived in a 
systematic way. When the energy is sufficiently low as for the fusion process considered here, the
effects of pions can be integrated out and absorbed into the coupling constants of
the contact interactions. The resulting effective field theory which is sometimes called 
EFT($\pi\!\!\! /)$ then involves only 
nucleon fields\cite{CRS}. In proton-proton scattering at low energies the Coulomb repulsion has a
dominant role and can naturally be incorporated into this theory\cite{KR}. As a direct result one 
can derive the difference between the strong scattering length which should be approximately the same
as in proton-neutron scattering, and the observed one which is modified by Coulomb effects. The relation
is very similar to the old result by Jackson and Blatt\cite{BJ}. Corrections due to effective-range
interactions can also be included but with more difficulty due to highly divergent integrals
involving Coulomb wavefunctions\cite{KR_pp}.

With this understanding of low-energy proton-proton elastic scattering, one can calculate the leading
order result for the fusion process $p + p \ra d + e^+ + \nu_e$ taking place at essential zero
initial kinetic energy\cite{KR_fus1}. With the use of a non-standard representation of the Coulomb propagator,
the result is in full agreement with the corresponding leading order nuclear physics result and depends
only on the physical proton-proton scattering length. To next order in the effective field theory 
expansion, one can derive higher order corrections\cite{KR_fus2} which also have the same structure as 
the corresponding effective-range corrections from more standard nuclear physics\cite{ES}\cite{BM}. 
However, at this order there appears an unknown counterterm in the effective Lagrangian which will enter 
as a correspondingly unknown term in the result for the fusion
rate. It can be determined from other related reactions. The most promising is neutrino scattering
on deuterons at low energies which has been investigated in the same effective theory by Butler and
Chen\cite{BC}.
When this is done, we will also have a more accurate and predictive result for the fusion rate.

In the next section we present the theoretical framework which in the following will be used to calculate
the proton-proton fusion rate in next-to-leading order in the momentum expansion of the effective theory.
This is done both for the proton-neutron and proton-proton sectors. A short summary of the leading
order calculation is given in section 3 followed by a more detailed calculation of the next-to-leading order
corrections. The derivation of the rate is completed by the inclusion of the effects of the 
counterterm. In the last section the obtained result is discussed and compared with what is obtained
by other methods. Using a recently improved 
matching of the coupling constants in the proton-neutron sector\cite{PRS}, we obtain a final result of
$\Lambda^2(0) = 7.37$ for the squared reduced matrix element giving the fusion rate. This is to be 
compared with the result of $\Lambda^2(0) = 7.08$ derived within the corresponding effective range 
approximation of nuclear physics. From the dependence
of the result on the magnitude of the unknown counterterm, we estimate the uncertainty to 
be 6\% - 8\%. This is significantly more than in other approaches where the unknown counterterm is
replaced by definite mesonic contributions with a resulting accuracy claimed to be close to 1\%. In
the present effective theory a corresponding accuracy can only be hoped for when the counterterm is
accurately determined in some other process. Finally, in an appendix we present a new and simpler method to
regularize divergent integrals involving derivatives of the Coulomb wavefunctions at the origin.

\section{Theoretical framework}
In the fusion reaction $p + p \ra d + e^+ + \nu_e$ at low energies the incoming protons
are in an antisymmetric spin singlet state. The deuteron $d$ has spin $S=1$ and the process
is thus a Gamow-Teller transition mediated by the weak axial current operator ${\bf A}_{-}$ which also
lowers the isospin by a unit. For a given kinetic energy $E$ and relative velocty $v_{rel}$ of 
the protons in the initial state, the reaction cross-section  is then given by the standard formula
\beq
    \sigma(E) = {G_A^2 m_e^5\over 2\pi^3 v_{rel}}\,f(E)\,|\bra{d}{\bf A}_{-}\ket{pp}|^2
                                                                              \label{rate}
\eeq
where the squared matrix element must be spin averaged. Here $G_A$ is the weak axial vector 
coupling constant, $m_e$ is the electron mass and $f(E)$ is the Fermi function resulting from the
integration over the available phase space of the leptons in the final state\cite{Bowler}. The available 
energy in the process for fusion at rest is set by the neutron-proton mass difference 
$\Delta M = M_n - M_p$ which is 1.294 MeV and the deuteron binding energy $B$ = 2.225 MeV. 
This gives an energy release of $0.931$ MeV carried away by the leptons. The temperature in the 
core of the Sun is approximately $15\times 10^6$\,K which corresponds to an average proton 
momentum around $p = 1.5$\, MeV and a much smaller energy. We will therefore in the following
assume that the initial proton energy $E$ approaches zero. The kinetic energy of the lepton pair 
will be much smaller than the momentum $\gamma = \sqrt{BM}$ of the bound nucleons with reduced 
mass $M/2$ in the deuteron. With the above value for the binding energy it follows that
$\gamma = 45.7$ MeV and thus to a very good approximation one can just ignore the momentum 
transfer between the leptons and the nucleons. 

The difficult part of calculating the fusion cross-section (\ref{rate}) lies in the hadronic
matrix element $T_{fi}(p) = \bra{d}{\bf A}_{-}\ket{pp}$ which is a function of the initial 
proton momentum $p = \sqrt {EM}$. Its magnitude can easily be estimated\cite{Bowler}. When the 
proton momentum goes to zero, the $pp$ wavefunction $\psi_p(r)$ becomes constant over the 
range of the deuteron. It is simply given by the the Sommerfeld factor $C_\eta = 
e^{-\pi\eta/2}|\Gamma(1 + i\eta)|$ where $\eta = \alpha M/2p$ characterizes the strength of the 
Coulomb repulsion between the protons\cite{LL}. The probability to find the two protons at the
same point is therefore
\beq
     C_\eta^2 = {2\pi\eta\over e^{2\pi\eta} - 1}                          \label{Sommer}
\eeq
At very low energies when $\eta$ gets large, it becomes exponentially small and is the
dominant effect in the fusion reaction. Similarly, in lowest order the deuteron wavefunction is simply
\beq
     \psi_d(r) = \sqrt{\gamma\over 2\pi}\;{e^{-\gamma r}\over r}        \label{psid}
\eeq
A rough estimate for the nuclear matrix element is then
\beq
      T_{fi}(p) = \int\!d^3r\,\psi_d(r)\psi_{p}(r) = C_\eta \sqrt{8\pi\over\gamma^3}
\eeq
This result sets the scale for the fusion rate. It is therefore natural to define the reduced 
matrix element\cite{ES}
\beq
    \Lambda(p) = \sqrt{{\gamma^3\over 8\pi C_{\eta}^2}} \, |T_{fi}(p)|    \label{Lambda}
\eeq
which contains all the interesting and important physics. It is expected to have a value of
the order of one and is now the conventional way of presenting theoretical results for the fusion
rate. The goal of the present paper is to calculate this number in a more model-independent way by 
purely analytical methods without using any other phenomenological input than the scattering lengths 
and effective ranges appearing in nucleon-nucleon scattering.

\subsection{KSW effective field theory}
During the last couple of years much progress has been made in understanding the low-energy 
properties of few-nucleon systems from the non-relativistic effective field theory proposed by Kaplan, 
Savage and Wise\cite{KSW_1}. At these length scales the proton and neutron are 
considered to be structureless point particles described by a nucleon isodoublet  $N^T = (p,n)$ 
Schr\"odinger field. For energies well below the pion mass $m_\pi$, all interactions including 
those due 
to pion exchanges, will be local. In the effective Lagrangian they can be thus represented by
terms involving only the nucleon field and derivatives thereof in such a way that all the 
symmetries obeyed by the strong interactions are preserved. At the
lowest energies only $S$-waves will contribute. Including no more than terms of dimension eight
in the derivative expansion, there are only two possible interaction terms in the Lagrangian 
parametrized by the coupling constants $C_0$ and $C_2$. It can be written as
\beq  
     {\cal L}_0 &=& N^\dagger\left(i\del_t + {\nabla^2\over 2M}\right)N
              - C_0(N^T{\bg\Pi}N)\cdot(N^T{\bg\Pi}N)^\dagger  \nn \\
    &+& {1\over 2}C_2\left\{(N^T\anabla^2{\bg\Pi}N)\cdot(N^T{\bg\Pi}N)^\dagger + h.c.\right\}
                                                                 \label{Leff}       
\eeq
where the operator $\anabla = (\overrightarrow{\nabla} - \overleftarrow{\nabla})/2$.
The projection operators $\Pi_i$ enforce the correct spin and 
isospin quantum numbers in the channels under investigation. More specifically, for 
spin-singlet interactions $\Pi_i = \sigma_2\tau_2\tau_i/\sqrt{8}$ while for spin-triplet 
interactions $\Pi_i = \sigma_2\sigma_i\tau_2/\sqrt{8}$. This theory is now valid below 
an upper momentum $\Lambda$ which will be the physical cutoff when the theory is regularized 
that way. Since the pion field is integrated out, all its effects are soaked up in the two 
coupling constants $C_0$ and $C_2$. Then the value of the cutoff $\Lambda$ will be set by the 
pion mass $m_\pi$. In this momentum range 
all the main properties of few-nucleon systems are now in principle given by the above
Lagrangian. More accurate results will follow from higher order operators in this 
field-theoretic description\cite{CRS}.  

The effective Lagrangian (\ref{Leff}) is non-renormalizable and divergent loop integrals
must be regularized. For this purpose one can use the OS scheme of Mehen and
Stewart\cite{OS} which is a generalization of the original proposal by Gegelia\cite{Gegelia}.
An equivalent method is the PDS scheme which was invented  by Kaplan, Savage and 
Wise\cite{KSW_1}\cite{KSW_2} and is based on dimensional regularization. We will use it here.
The {\it a priori} 
unknown coupling constants $C_0$ and $C_2$ can be determined in terms of experimental 
quantities measured in low-energy nuclear reactions. The size of the dimension-six coupling 
constant $C_0$ will then be determined by the scattering length $a$ in nucleon-nucleon scattering
while $C_2$ is found to be proportional to the effective range parameter $r_0$. Both of these
coupling constants will depend on the renormalization mass $\mu$ which enters in the PDS
regularization scheme. It can be chosen freely in the interval $\gamma < \mu \le m_\pi$ but 
physical results obtained from the effctive theory should be independent of its precise value.

\subsection{Proton-neutron interactions and the deuteron}
The deuteron will appear as a bound  state in proton-neutron scattering in the spin-triplet channel.  
It is then natural to determine the corresponding coupling constants by matching the 
results to properties at the deuteron pole of the scattering amplitude. The residue at the pole
gives the renormalization constant $Z$ of the deuteron interpolating field which replaces the
wavefunction of the bound  state\cite{KSW_2}.
\begin{figure}[ht]
   \begin{center}
          \epsfig{figure=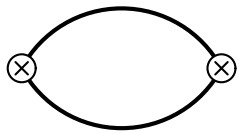,height=17mm}
   \end{center}
   \vspace{-4mm}            
   \caption{\small Feynman diagram representing the leading order contribution to the deuteron
   wave\-function renormalization. The crosses represent the coupling to the interpolating field for the
   deuteron while the lines are nucleon propagators.}
   \label{fig.1}
\end{figure}
In lowest order of perturbation theory one finds the renormalization constant from the 
irreducible 2-point function $\Sigma(E)$ shown in Fig. 1. At the two vertices the 
interpolating field acts with energy $E$ and zero momentum and a strength which we choose to be 
$-1$. In the intermediate state there is a 
neutron and a proton which propagate with relative momentum $\bk$. Integrating over all these
momenta we then find the value of the diagram,
\be
     \Sigma_0(E) = \int\!{d^3k\over(2\pi)^3} {1\over E - \bk^2/M + i\e}          \label{Sigma}
\ee
This divergent integral is now made finite using the PDS regularization scheme\cite{KSW_1}
and gives
\beq
    \Sigma_0(E) = -{M\over 4\pi}\left(\mu - \sqrt{-ME}\right)         \label{Sigma_0}
\eeq
The renormalization constant 
\beq
         Z = \sqrt{ 1\over |d\Sigma/dE|}_{E= - B}            \label{Z_0}
\eeq
which is evaluated
at the deuteron pole where the binding energy is $B = \gamma^2/M$, thus takes the value
$Z_0 = \sqrt{8\pi\gamma}/M$ at this order of perturbation theory. It is independent of the
coupling constant $C_0^d$ whose effect must be summed to all orders in order to find the
non-perturbatively bound state in this channel. When it takes the special renormalized value
\beq
     C_0^d(\mu) = {4\pi\over M} {1\over\gamma - \mu}                      \label{C0d}
\eeq
we see that the reducible chain of bubble interactions in Fig. 2 just gives the same result
as for the single bubble in the irreducible diagram in Fig. 1. 
\begin{figure}[ht]
   \begin{center}
          \epsfig{figure=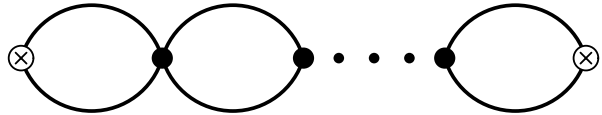,height=15mm}
   \end{center}
   \vspace{-4mm}            
   \caption{\small Chain of proton-neutron interactions mediated by the leading order contact term.}
   \label{fig.2}
\end{figure}
In this particular channel one shall therefore not sum such chains of bubble diagrams when one describes 
the bound state deuteron by an interpolating quantum field.

While the coupling constant $C_0^d$ must be treated non-perturbatively, the effects of the derivative
coupling $C_2^d$ are included only to first order. The corresponding renormalized coupling constant is
found to be $C_2^d(\mu) = \rho_d M(C_0^d(\mu))^2/8\pi$ where $\rho_d = 1.76$ fm is the spin-triplet 
$pn$ effective range scattering parameter evaluated at the deuteron pole\cite{KSW_2}\cite{CRS}. 
It will also contribute to the renormalization constant $Z$
via the perturbative diagram in Fig. 3 for the 2-point function. It has the value $\Sigma_2(E) =
C_2^d ME \Sigma_0^2(E)$ which gives the total contribution
\beq
    Z_2 = Z_0
    \left[1 - {\gamma M\over 2\pi}C_2^d(\mu - \gamma)(\mu - 2\gamma)\right]^{-1/2} \label{Z2}
\eeq
In the limit $\mu \gg \gamma$ the dependence on the regularization mass $\mu$ is seen to go away.
The previous value $Z_0$ then gets modified by the factor $\sqrt{Z_d} = 1/\sqrt{1-\gamma\rho_d}$. 
This corresponds to a change of the normalization of the deuteron wavefunction (\ref{psid}) which now 
becomes $\psi_d(r) \ra \sqrt{Z_d}\psi_d(r)$ in agreement with effective range theory in nuclear 
physics\cite{effBethe}. 
\begin{figure}[ht]
   \begin{center}
          \epsfig{figure=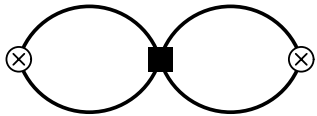,height=15mm}
   \end{center}
   \vspace{-4mm}            
   \caption{\small Effective-range correction to the deuteron wavefunction renormalization constant.}
   \label{fig.3}
\end{figure}
Here it is only valid at large distances since properties of the deuteron at scales less than $1/m_\pi$
are not accessible in this theory. Also, it is strictly only valid to first 
order in an expansion in powers of $\gamma\rho_d$ since 
it is obtained perturbatively in the coupling constant $C_2^d$. Since the expansion
parameter has the rather large value $\gamma\rho_d =0.41$, it is desirable to improve the 
convergence of perturbation theory in this coupling constant. This has recently been achieved by 
Phillips, Rupak and Savage\cite{PRS} whose method we will apply at the end of the more
conventional approach we present first.

\subsection{Coulomb interactions and the proton-proton wavefunction}
In the absence of strong interactions, the incoming proton-proton state with center-of-mass
momentum $\bp$ is given by the Coulomb wavefunction\cite{AS} 
\beq
   \psi_\bp(\br) = {1\over\rho}\sum_{\ell = 0}^\infty (2\ell + 1)i^\ell e^{i\sigma_\ell} 
                   F_\ell(\rho)P_\ell(\cos(\theta)                         \label{psip}
\eeq
Here $\rho = pr$ and $\sigma_\ell = \arg{\Gamma(1+\ell + i\eta)}$ is the Coulomb phaseshift. 
At low energies only the $S$-wave will contribute. It is given in terms of the Kummer 
function $M(a,b;z)$ as
\beq
    F_0(\rho) = C_\eta \rho e^{-i\rho}M(1-i\eta,2;2i\rho)                   \label{F0}
\eeq
which is a confluent hypergeometric function.

The strong interactions between the protons can now be included  using the same KSW Lagrangian 
(\ref{Leff}) but now with coupling constants $C_0^p$ and $C_2^p$ which also get 
renormalized\cite{KR_pp}. 
\begin{figure}[ht]
   \begin{center}
          \epsfig{figure=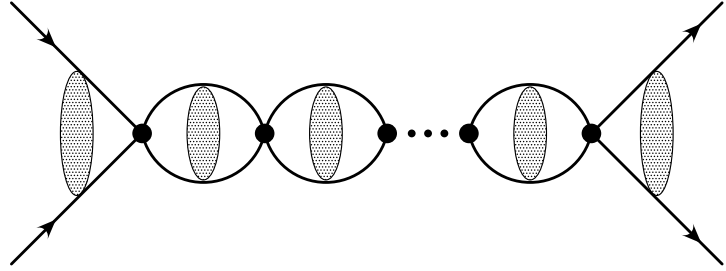,height=35mm}
   \end{center}
   \vspace{-4mm}            
   \caption{\small Elastic scattering due to chain of bubble diagrams with Coulomb interactions. 
                   Incoming and outgoing particles are in Coulomb eigenstates. }
   \label{fig.4}
\end{figure}
As in the proton-neutron channel one must again consider the coupling
$C_0^p$ to all orders in perturbation theory. In this way one finds that proton-proton elastic
scattering is given by the infinite sum of all chains of Coulomb-dressed bubble diagrams
as shown in Fig. 4. Each bubble is given by the Coulomb propagator
\beq
      G_C(E;\br',\br) =  M\!\int\!{d^3 q\over (2\pi)^3}
     {\psi_\bq(\br')\psi_\bq^*(\br)\over p^2 -q^2 + i\e}                  \label{GC1}
\eeq
It satisfies the Lippmann-Schwinger equation  $G_C = G_0 + G_0V_CG_C$ where $V_C$
is the Coulomb potential and 
\beq
       G_0(E;\bq',\bq) =  {M\over p^2 - q^2 + i\e}(2\pi)^3\delta(\bq' - \bq)       \label{G0}
\eeq
is the free propagator in momentum space. Iterating this functional equation we see from Fig. 5
that it corresponds to the exchange of zero, one, two and more static photons.
Since a single bubble in Fig. 4 corresponds to the propagation of the proton pair with energy 
$E = p^2/M$ from zero separation and back to zero separation, it has the value
$J_0(p) =  G_C(E;\br'=0,\br=0)$ or
\beq
      J_0(p) = M\!\int\!{d^3 q\over (2\pi)^3} {2\pi\eta(q)\over e^{2\pi\eta(q)} - 1}
               {1\over p^2 - q^2 + i\e}                                      \label{JC}
\eeq
\begin{figure}[ht]
   \begin{center}
          \epsfig{figure=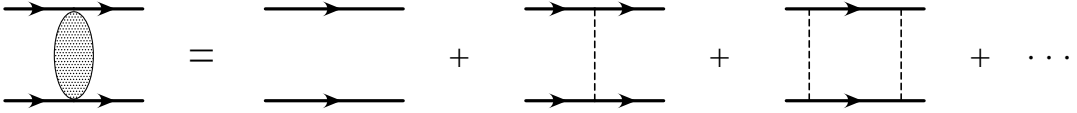,height=15mm}
   \end{center}
   \vspace{-4mm}            
   \caption{\small The Coulomb propagator can be represented by an infinite sum of exchanged 
                   static photons between the two charged particles.}
   \label{fig.5}
\end{figure}The integral is seen to be ultraviolet divergent, but can be regularized in the PDS scheme in 
$d = 3 - \e$ dimensions. When
contributions from poles in $d=2$ dimensions are subtracted, one finds\cite{KR}
\beq
    J_0(p) = {\alpha M^2\over 4\pi}\left[{1\over\e} + \ln{\mu\sqrt{\pi}\over\alpha M} + 1
             - {3\over 2}C_E - H(\eta) \right] - {\mu M\over 4\pi}           \label{J01}
\eeq
Here $C_E = 0.5772$ is Euler's constant and the function
\beq
    H(\eta) = \psi(i\eta) + {1\over 2i\eta} -\ln(i\eta)                         \label{Heta}
\eeq
is known to appear in these Coulomb scattering problems\cite{BJ}. The divergent $1/\e$ piece 
will be absorbed in counterterms representing electromagnetic
interactions at shorter scales. This replaces the bare coupling constant $C_0^p$ with the
renormalized value $C_0^p(\mu)$. It can be found by matching the calculated proton-proton 
scattering amplitude to the experimental one. This is usually given by the measured scattering 
length $a_p = -7.82$ fm when the proton momentum  $p \ra 0$.  Thus one finds\cite{KR}  
\beq
     {1\over C_0^p(\mu)} = {M\over 4\pi a_p} + J_0(0)                              \label{C0p}
\eeq     
Since the function (\ref{Heta}) is dominated by its real part $h(\eta) = 1/(12\eta^2) 
+ {\cal O}(\eta^{-4})$ which goes to zero when  $p \ra 0$, we see from the form of $J_0(0)$ in
(\ref{J01}) that it is natural to introduce the $\mu$-dependent scattering length
\beq
    {1\over a(\mu)} = {1\over a_p}  +  \alpha M
    \left[\ln{\mu\sqrt{\pi}\over\alpha M} + 1 - {3\over 2}C_E\right]             \label{Capp}
\eeq
where $\alpha$ is the fine-structure constant. It corresponds to the Jackson-Blatt relation 
between the strong and Coulomb-modified proton-proton scattering lengths\cite{BJ}. Then we can
write 
\beq
      C_0^p(\mu) = {4\pi\over M}{1\over 1/a(\mu) - \mu}                      \label{C0pmu}   
\eeq
which is now on the same form as (\ref{C0d}) for the bound-state case.

In next order of the momentum expansion the derivative coupling $C_2^p$ in (\ref{Leff}) is 
introduced perturbatively. Again matching to low-energy proton-proton scattering, one finds 
$C_2^p(\mu) = \rho_d M(C_0^p(\mu))^2/8\pi$ where $\rho_p = 2.79$ fm is the the 
proton-proton effective range parameter. It is not 
affected by Coulomb corrections to this order in the effective theory. However,
the $C_2^p$ coupling gives an important contribution to the scattering length (\ref{Capp})
which picks up an additional term $-\mu\rho_p/2$ in the parenthesis\cite{KR_pp}.

\subsection{Gamow-Teller transition operators}
The dominant weak transition matrix elements in the basic fusion rate formula (\ref{rate}) are 
due to the elementary isovector axial current operator 
\beq
        {\bf A}_{-}^{(1)} = N^\dagger{\bg\sigma}\tau_{-}N                     \label{A1}
\eeq
which converts an incoming proton into a neutron with the proper spin and isospin quantum
numbers. This is the ordinary one-body interaction depicted in Fig.6a. 
But when we include the dimension-eight derivative operator in (\ref{Leff}) higher
dimension weak transition operators must also be considered. These were first discussed
by Butler and Chen in connection with elastic and inelastic scattering of neutrinos on 
deuterons\cite{BC}. In our case there is only one such operator which can be written as
\beq
        {\bf A}_{-}^{(2)} = L_{1A}(N^T{\bg\Pi}N)^\dagger(N^T\Pi_{-}N)               \label{A_2}
\eeq
where the projection operator ${\bg\Pi} = \sigma_2{\bg \sigma}\tau_2/\sqrt{8}$
acts in the spin-triplet final proton-neutron state while $\Pi_{-} = 
\sigma_2\tau_2\tau_{-}/\sqrt{8}$ acts on the spin-singlet proton-proton initial state. 
The weak axial
\begin{figure}[ht]
   \begin{center}
          \epsfig{figure=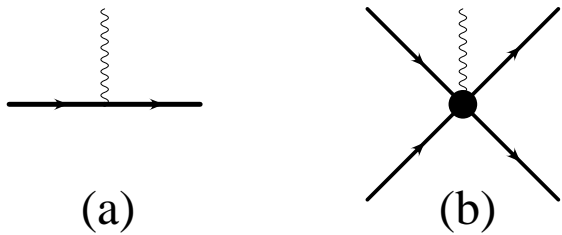,height=30mm}
   \end{center}
   \vspace{-4mm}            
   \caption{\small One-body interaction in (a) represents the weak axial current vertex while the
                   two-body interaction in (b) represents the higher order axial vector counterterm.}
   \label{fig.6}
\end{figure}
vector coupling constant has been factored out so that the effective coupling constant is
just $L_{1A}$. This new two-body operator represents weak transitions taking place at shorter length scales
than considered in the effective theory and the corresponding vertex is shown in Fig.6b. 
Typically it represents transitions due to pion 
interactions and other two-body interactions. In the effective theory it will act as a 
counterterm which can absorb the dependence on the renormalization mass $\mu$. Its
actual magnitude is presently unknown. It can be estimated from dimensional arguments combined
with the renormalization group. Even better would be to determine it in some other weak process
where its contribution could be isolated and measured.

\section{Hadronic matrix elements}
We are now in position to calculate the hadronic matrix elements 
$T_{fi}(p) = \bra{d}{\bf A}_{-}\ket{pp}$ of the weak transition operator. The initial proton 
state is constructed in terms of the Coulomb wavefunctions as in the elastic scattering
case. For the final state deuteron we use the interpolating field which contains a proton-neutron
state with the amplitude $Z$ which is the wavefunction renormalization constant. We will
initially consider only the action of the axial current operator (\ref{A1}). The effects of
spin has been separated out and will not enter the following calculation. 

\subsection{Leading order result}
In lowest order of the effective theory only the dimension-six operators will contribute with
coupling constants $C_0^d$ and $C_0^p$ in the deuteron and proton-proton sectors respectively.
\begin{figure}[ht]
   \begin{center}
          \epsfig{figure=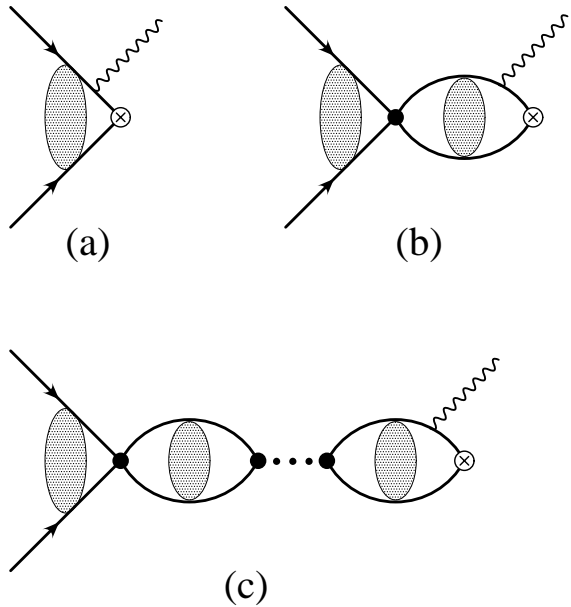,height=80mm}
   \end{center}
   \vspace{-4mm}            
   \caption{\small Feynman diagrams contributing to proton-proton fusion in leading order.}
   \label{fig.7}
\end{figure}
The transition matrix element $T_{fi}$ then gets contributions from three classes of diagrams 
shown in Fig. 7. After being hit by the weak current, the proton-proton system is transformed 
into a bound deuteron. The value of the simplest diagram in Fig.7a is then
seen to be $Z_0 A_0(p)$ where $Z_0$ is the  constant derived in the previous section and
\beq
      A_0(p)  = \int\!{d^3k\over(2\pi)^3} 
           {M\over \bk^2 + \gamma^2}\psi_\bp(\bk)                            \label{A0}
\eeq
There is a factor (-1) from the deuteron vertex and the bound proton-neutron propagator is 
$-M/(\bk^2 + \gamma^2)$. In addition, we have introduced the Fourier transform $\psi_\bp(\bk)$ 
of the Coulomb wavefunction (\ref{psip}) when the protons have the center-of-mass momentum $\bp$. 
Including next the strong interaction once between the two protons as shown in Fig.7b, we get the 
contribution $Z_0C_0B_0(p)\psi_\bp(0)$ where
\beq
    B_0(p)  =  \int\!{d^3k\over(2\pi)^3}{d^3k'\over(2\pi)^3} 
           {M\over \bk^2 + \gamma^2}G_C(E;\bk,\bk')                              \label{B0}
\eeq
is a convergent integral and the last factor $\psi_\bp(0) = C_\eta e^{i\sigma_0}$ gives the
amplitude for the two incoming protons to meet at the first vertex. Going to higher orders
in the coupling $C_0^p$ we will add in Coulomb-dressed bubble diagrams as in Fig.7c. Each bubble
is of the same form as in proton-proton scattering in Fig. 4 where the
contribution from each bubble is given by $J_0(p)$ in (\ref{JC}). Adding up these diagrams, they are seen
to form a geometric series with the sum $C_0^p/(1 - C_0^pJ_0)$. The total contribution
from all the three classes of diagrams thus gives the lowest order transition amplitude 
$T_{fi}(p) = Z_0\, T_0(p)$ where 
\beq
 T_0(p) =  \left[A_0(p) + B_0(p){C_0^p\psi_\bp(0)\over 1 - C_0^pJ_0(p)}\right]  \label{Tfi}
\eeq 
The term involving $C_0^p$ can now be expressed in terms of the proton-proton scattering
length $a_p$ in (\ref{C0p}) and is independent of the renormalization scale $\mu$. 

For the explicit evaluation of this matrix element it is necessary to introduce the  Coulomb 
wavefunction (\ref{F0}). Since the first term of the momentum integral is the product of two
Fourier transformed functions, we find that it simplifies in coordinate space to
\beq
     A_0(p) &=& M C_\eta e^{i\sigma_0}\int_0^\infty\!dr r e^{-(\gamma +ip)r}
             M(1-i\eta,2;2ipr) \\ \nn
          &=&   {M C_\eta e^{i\sigma_0}\over (\gamma + ip)^{2}}
                 {_2F_1}\left(1-i\eta,2;2;{2ip\over\gamma + ip}\right)
\eeq
Now the hypergeometric function $_2F_1(a,b,b;z) = (1 - z)^{-a}$ so that the final result 
can be written as
\beq
    A_0(p) = C_{\eta}e^{i\sigma_0}{M\over p^2 +\gamma^2}
            e^{2\eta\arctan({p\over\gamma})}                                \label{Ap}
\eeq
In the expression (\ref{B0}) for $B_0(p)$ we notice that the integral over $\bk'$ gives 
the complex conjugate value of the Coulomb wavefunction at the origin. It therefore takes 
the form
\be
    B_0(p) =   M \int\!{d^3k\over(2\pi)^3}\int\!{d^3q\over(2\pi)^3} 
            {M\over \bk^2 + \gamma^2}{\psi_\bq(\bk)\over\bp^2 -\bq^2 + i\e} 
           \psi_\bq^*(0)     
\ee
The integral over $\bk$ is just the previous result for $A_0(q)$ so that
\beq
    B_0(p) = M \int\!{d^3q\over(2\pi)^3}{M\over q^2 +\gamma^2} 
            {e^{2\eta\arctan({q\over\gamma})}\over p^2 - q^2 + i\e}
            {2\pi\eta(q)\over e^{2\pi\eta(q)} - 1}                            \label{Bp}
\eeq
When the momentum of the incoming proton is non-zero it yields in general a complex result.

In the fusion limit $p\ra 0$ we now find that the first term (\ref{Ap}) simplifies to
\beq
    A_0(p\ra 0)  = C_\eta\, {M\over\gamma^2} \, e^{\chi + i\sigma_0}
\eeq
where the parameter $\chi =\alpha M/\gamma$. Similarly, the second term $B_0(p)$  becomes 
proportional to the integral 
\beq
    I(\chi) = \int_0^\infty\!dx {2x\over e^x -1}
    {e^{{x\over\pi}\arctan({\pi\chi\over x})}\over x^2 + \pi^2\chi^2}      \label{Int}
\eeq
in the same limit when we use $x = 2\pi \eta(q)$ as a new integration variable. Repeating this 
calculation with a different representation of the Coulomb Green's function, it can be shown
that the integral takes the value\cite{KR_fus1}
\beq
     I(\chi)= {1\over\chi} - e^\chi E_1(\chi)
\eeq
when expressed in terms of the exponential integral
\be
    E_1(\chi) = \int_\chi^\infty dt {e^{-t}\over t}
\ee
With $Z_0 = \sqrt{8\pi\gamma}/M$ for the renormalization constant, we thus find for the full 
matrix element the result  
\beq
     T_{fi} = \sqrt{8\pi\over \gamma^3}\,C_\eta\, e^{i\sigma_0}
              \left[ e^\chi - \alpha M a_p\,I(\chi)\right]
\eeq
The reduced matrix element in leading order is therefore
\beq
    \Lambda_0(0) = e^\chi - \alpha M a_p\,I(\chi)                   \label{L0}
\eeq
This is also the canonical result from standard nuclear physics\cite{BM}. The parameter
$\chi = 0.15$ and thus the integral $I(0.15)= 4.96$. Combined with the measured 
value $a_p = -7.82$\,fm for the scattering length, we then have $\Lambda_0(0) = 2.51$ for the
reduced matrix element. In the formula for the fusion rate it gives the contribution 
$\Lambda_0^2(0) = 6.30$. From previous applications of the effective theory\cite{EFTbook}, we know 
that leading-order results are typically within 20 - 30\% of the correct values. Going to next order 
in perturbation theory, the accuracy is expected to increase to 5 - 10\%.

\subsection{Effective range corrections}
In next order of the momentum expansion of the effective field theory, there is no operator
which induces $S - D$ mixing of the deuteron state. It will first appear at one order higher\cite{KSW_2}. 
\begin{figure}[ht]
   \begin{center}
          \epsfig{figure=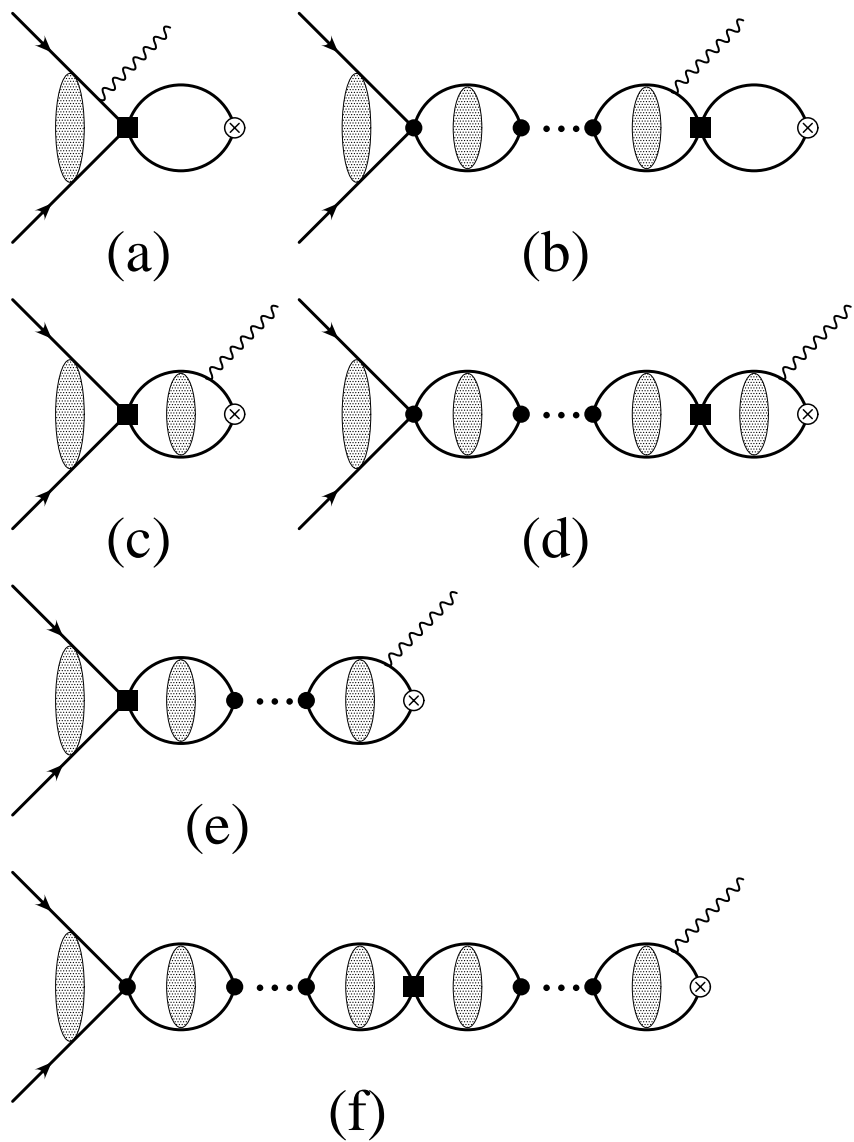,height=150mm}
   \end{center}
   \vspace{-4mm}            
   \caption{\small Corrections to the fusion amplitude coming in at next-to-leading order.}
   \label{fig.8}
\end{figure}
The dimension-eight couplings $C_2^d$ and $C_2^p$ give the additional diagrams shown 
in Fig. 8 in first order perturbation theory. Each such operator $V_2$ has a momentum matrix element 
$\bra{\bk}V_2\ket{\bq} = C_2(\bk^2 + \bq^2)/2$. The contribution from Fig.8a is seen to be
\beq
       T_{a} = {1\over 2}C_2^d \int\!{d^3k\over(2\pi)^3} \int\!{d^3q\over(2\pi)^3} 
{-M\over \bk^2 + \gamma^2}(\bq^2 + \bk^2){M\over \bq^2 + \gamma^2}\psi_\bp(\bq)  \label{Ta1}
\eeq    
This can be expressed in terms of the divergent integral
\beq
       I_0(\gamma) =  -M\int\!{d^3k\over(2\pi)^3}{1\over \bk^2 + \gamma^2} 
           = {-M\over 4\pi}(\mu - \gamma)                                       \label{I0}
\eeq
which is the same as occured in the lowest-order determination of the wavefunction
renormalization constant in (\ref{Sigma_0}). It is finite after PDS regularization which gives
for the other occuring integral
\beq
       I_2(\gamma) =  -M\int\!{d^3k\over(2\pi)^3}{\bk^2\over \bk^2 + \gamma^2} 
           = -\gamma^2 I_0(\gamma)                                              \label{I2}
\eeq
since $\int d^dk/(2\pi)^d = 0$ in dimensional regularization. Together with the function $A_0(p)$ 
in (\ref{A0}) and the related function
\beq
      A_2(p)  = M\int\!{d^3k\over(2\pi)^3} 
           {\bk^2\over \bk^2 + \gamma^2}\psi_\bp(\bk)                            \label{A2}
\eeq
we thus have for the matrix element (\ref{Ta1})
\be
     T_{a} = {1\over 2}C_2^d\left[I_2(\gamma) A_0(p) + I_0(\gamma) A_2(p)\right]
\ee
In the same way as we could express the integral $I_2(\gamma)$ in terms of $I_0(\gamma)$, 
we also find
\beq
      A_2(p) = M\psi_\bp(0) - \gamma^2A_0(p)                            \label{A2a}
\eeq
Here $\psi_\bp(0) = C_\eta e^{i\sigma_0}$ where $\sigma_0$ is the Coulomb $S$-wave phaseshift. When 
we eventually use this result to calculate the fusion rate from (\ref{Lambda}), we will take 
the absolute value and this phase factor will not contribute. We therefore write
\be
     T_{a} = {1\over 2}C_2^d I_0(\gamma)\left[MC_\eta(p) - 2\gamma^2 A_0(p)\right]
\ee
where the same phase factor also should be dropped in the last term. This result is now to be 
taken in the fusion limit $p\ra 0$ as in the previous section.

The contribution from the diagram Fig.8b involves the Coulomb Green's function and its
derivative in the triple integral
\be T_{b1} =
{1\over 2}C_2^d C_0^p\int\!{d^3k\over(2\pi)^3} \int\!{d^3q\over(2\pi)^3}\int\!{d^3q'\over(2\pi)^3}
 G_C(E;\bq,\bq'){M\over \bq^2 + \gamma^2}(\bq^2 + \bk^2){-M\over \bk^2 + \gamma^2}\psi_\bp(0)
\ee
resulting from just one proton bubble. This can be expressed in terms of the function $B_0(p)$ in 
(\ref{B0}) and the related function
\beq
    B_2(p)  = M \int\!{d^3k\over(2\pi)^3}{d^3q\over(2\pi)^3} 
           {\bk^2\over \bk^2 + \gamma^2}G_C(E;\bk,\bq)                              \label{B2}
\eeq
as 
\be
    T_{b1} =  {1\over 2}C_2^d C_0^p [I_2(\gamma) B_0(p) + I_0(\gamma)B_2(p)]\psi_\bp(0)
\ee
With the simplification
\beq
    B_2(p) = MJ_0(p) - \gamma^2B_0(p)                                            \label{B2a}
\eeq 
we find the total contribution
\beq
     T_{b} = {1\over 2}C_2^d {I_0(\gamma) C_0^p\over 1 - C_0^pJ_0(p)}\left[M J_0(p) 
           - 2\gamma^2B_0(p)\right]C_\eta
\eeq
from all the Coulomb-dressed proton bubble diagrams in Fig.8b. Here we have again replaced 
$\psi_\bp(0)$ by $C_\eta(p)$.

The remaining diagrams involve the proton derivative coupling $C_2^p$. Diagram Fig.8c gives
\beq T_c =
{1\over 2}C_2^p \int\!{d^3q\over(2\pi)^3} \int\!{d^3k\over(2\pi)^3}\int\!{d^3k'\over(2\pi)^3}
        {M\over \bk'^2 + \gamma^2} G_C(E;\bk',\bk)(\bk^2 + \bq^2)\psi_\bp(\bq)
\eeq
This can again be expressed in terms of the functions  $B_0(p)$ and $C_\eta(p)$ 
and their derivatives. In particular, we define
\beq
    B_2'(p) = M \int\!{d^3k\over(2\pi)^3}{d^3q\over(2\pi)^3} 
           {\bq^2\over \bk^2 + \gamma^2}G_C(E;\bk,\bq)                              \label{B2prim}
\eeq
and introduce 
\beq
    \psi_2(p) = \int\!{d^3k\over(2\pi)^3}\bk^2 \psi_\bp(\bk)                        \label{psi2}
\eeq
which is the double derivative of the Coulomb wavefunction at the origin. Both of them are 
highly divergent, but can be calculated in the PDS regularization scheme and expressed
in terms of already introduced functions. This is shown in the appendix. We thus find for this
diagram
\be
     T_c = {1\over 2}C_2^p\left[\psi_0(p)B_2'(p) + \psi_2(p)B_0(p)\right]         \label{Tc}
\ee
where $\psi_2(p) = -\alpha M\mu\psi_0(p)$ in the limit $p\ra 0$ as shown in the appendix. 
Then we also have $B_2'(p) = MI_0(\gamma) - \alpha M\mu B_0(p)$ and we therefore get
\beq
    T_c  =  {1\over 2}C_2^p C_\eta\left[MI_0(\gamma) - 2\alpha M\mu B_0(p)\right] \label{Tc1}
\eeq
which now involves only finite and known quantities.

In Fig.8d we sum over all the Coulomb-dressed proton bubbles. The result is given by the
multiple integral
\be 
    T_{d} &=& {1\over 2}C_2^p {C_0^p\over 1 - C_0^pJ_0(p)}
            \int\!{d^3k\over(2\pi)^3}\int\!{d^3k'\over(2\pi)^3} 
            \int\!{d^3q\over(2\pi)^3}\int\!{d^3q'\over(2\pi)^3}
            G_C(E;\bq,\bq')(\bq^2 + \bk^2)                     \\
          &\times& G_C(E;\bk',\bk){M\over \bk'^2 + \gamma^2}\psi_\bp(0)
\ee
Again we can reorder the integrand so that the result is expressed in terms of simpler functions,
\be 
   T_{d} &=& {1\over 2}C_2^p{C_\eta C_0^p\over 1 - C_0^pJ_0(p)}
             \left[J_2(p)B_0(p) + J_0(p)B_2'(p)\right]
\ee
where now
\beq
    J_2(p) = \int\!{d^3k\over(2\pi)^3}\int\!{d^3q\over(2\pi)^3}\bk^2G_C(E;\bk,\bq)     \label{J2p}
\eeq
involves the derivative of the Coulomb propagator. It is also evaluated in the appendix. 
In the limit $p\ra 0$ we find $J_2(p) = -\alpha M\mu J_0(p)$ which together with the related 
result for $B_2'(p)$  gives
\beq
     T_d = {1\over 2}C_2^p {C_\eta C_0^p J_0(p)\over 1 - C_0^pJ_0(p)}
            \left[MI_0(\gamma) - 2\alpha M\mu B_0(p)\right]
\eeq
It has the same structure as $T_c$ in (\ref{Tc1}) and they can therefore be combined into a simpler
result.

The deuteron side of the diagrams in Fig.8e is seen to be just $B_0(p)$. Summing up the
bubbles on the proton side, we find 
\be
    T_e = {1\over 2} {C_2^p C_0^p\over 1 - C_0^pJ_0(p)}
          \left[J_2(p)\psi_0(p) + J_0(p)\psi_2(p)\right]B_0(p)
\ee
In the limit  $p \ra 0$ this simplifies again with the result 
\beq
    T_e = -C_2^p C_\eta {C_0^p J_0(p)\over 1 - C_0^p J_0(p)} \alpha M\mu B_0(p)
\eeq
Similarly we find that the diagrams in Fig.8f gives
\be
    T_f = {1\over 2}C_2^p \left({C_0^p\over 1 - C_0^pJ_0(p)}\right)^2
          \left[J_2(p)J_0(p) + J_0(p)J_2(p)\right]\psi_0(p)B_0(p)
\ee
which becomes
\beq
    T_f = - C_2^p C_\eta \left({C_0^p J_0(p)\over 1 - C_0^pJ_0(p)}\right)^2
            \alpha M\mu B_0(p)
\eeq
in the low-energy limit $p\ra 0$. 

Adding now up the contributions from all the diagrams in Fig. 8, we obtain the sum 
\be
     T_2 &=& -  \gamma^2C_2^d I_0(\gamma) T_0(p) 
           + {1\over 2}(C_2^p + C_2^d){MI_0(\gamma) C_\eta\over 1 - C_0^pJ_0(p)} \\
     &-& {  \alpha M\mu C_2^p C_\eta B_0(p)\over (1 - C_0^pJ_0(p))^2} 
\ee 
where $T_0(p)$ is the lowest order matrix element (\ref{Tfi}). The full transition matrix element 
to this order is therefore
\be
    T_{fi} = Z_2 T_0(p) + Z_0 T_2(p)
\ee
where $Z_2$ is the next-to-leading order renormalization constant (\ref{Z2}). Reordering
and combining terms, we obtain
\beq
   Z_0^{-1}T_{fi} &=& A_0(p) + B_0(p)C_\eta\left[{ C_0^p\over 1 - C_0^pJ_0(p)} 
    - {\alpha M\mu\, C_2^p\over  [1 - C_0^pJ_0(p)]^2}\right] \\
    &+& {\gamma M\over 4\pi}C_2^d(\mu)(\mu - \gamma)^2\left[A_0(p) 
    + B_0(p){C_\eta C_0^p\over 1 - C_0^pJ_0(p)}\right] \\
   &-& {M^2\over 8\pi}C_\eta (\mu - \gamma)
   {C_2^p + C_2^d\over 1 - C_0^pJ_0(p)}           \label{Tfi2}
\eeq
In the bubble integral $J_0(p)$ we can take $p \ra 0$ since it is finite.
The function $B_0(p)$ is also finite in this limit while $A_0(p)$ becomes proportional to the Coulomb
factor $C_\eta(p)$ which diverges. As shown previously in the application of the same effective 
theory to low-energy,
elastic proton-proton scattering, the first square bracket is now just the physical proton-proton 
scattering length $a_p$ calculated in next-to-leading order with the result\cite{KR_pp}
\beq
     a_p = {M\over 4\pi}\left({ C_0^p\over 1 - C_0^pJ_0(0)} 
         -{\alpha M\mu\, C_2^p\over [1 - C_0^pJ_0(0)]^2}\right)              \label{ap2}
\eeq
The last term is the effective-range correction which is important in order to have a physically
meaningful result for the scattering length. We see that when this is zero, we have the  
previous result (\ref{C0p}) used in leading order.

The transition matrix element in next-to-leading order is now given by (\ref{Tfi2}). Isolating
a common factor, the reduced matrix element (\ref{Lambda}) follows as
\beq
    \Lambda_2(0) =  \Lambda_0(0)\left[1 + {\gamma M\over 4\pi}C_2^d(\mu)(\mu - \gamma)^2\right]
                 - a_p\gamma^2(\mu - \gamma) {C_2^p(\mu) + C_2^d(\mu)\over 2C_0^p(\mu)}   \label{L2}
\eeq
where $\Lambda_0(0)$ is the leading-order result (\ref{L0}) but now expressed in terms of the 
next-to-leading order scattering length (\ref{ap2}). With the already established value for the 
coupling constant $C_2^d$ we see that it is now multiplied by the factor $1 + \gamma \rho_d/2$. 
This can be interpreted as the first term in the expansion of the deuteron normalization
factor $Z_d$ discussed in the previous section. While this term in the result is independent of the 
renormalization scale $\mu$, we see that the last term is generally not. However, when $\mu$ is much larger
than the other mass scales given by the scattering lengths, this dependence goes away and we are
left with the definite result
\beq
    \Lambda_2(0)_{\mu\gg\gamma} =   \Lambda_0(0)\left[1 + {1\over 2}\gamma\rho_d\right]
                     + {1\over 4}a_p\gamma^2(\rho_p + \rho_d)              \label{NLOX}
\eeq
It has a structure which is very similar to the reduced matrix element in the standard nuclear physics 
effective-range approximation\cite{ES}\cite{BM}
\beq
   \Lambda_{ER}(0) = \sqrt{Z_d}\left[\Lambda_0(0) + {1\over 4}a_p\gamma^2(\rho_p + \rho_d)\right]\label{ER}
\eeq
With the known values for the different nucleon parameters, the result in this old approximation is therefore 
$\Lambda_{ER}(0) = 2.66$ or $\Lambda_{ER}^2(0) = 7.08$. On the other hand, our
next-to-leading order result (\ref{NLOX}) gives $\Lambda_2(0)_{\mu\gg\gamma} = 2.54$ which is just 
a 1.4\% addition to the leading
order result we previously obtained . This is surprisingly small, but results from an almost total
cancellation between the two effective-range corrections in (\ref{NLOX}). The net result for the
squared matrix element is $\Lambda_2^2(0)_{\mu\gg\gamma} = 6.45$ which is seen to be 8\% below the 
effective-range value.

\subsection{Contribution from counterterm}
A complete calculation of the fusion rate in next-to-leading order must include all operators
contributing to this order in the momentum expansion of effective theory. Until now we have
only included the effects of the dimension-eight operators coupling four nucleons with a derivative
interaction. Since our result above in general depends on the renormalization scale, it signals 
that the calculation is incomplete. There should be additional interaction terms that in principle
should absorb all dependence on the renormalization scale. This is in fact the case
as shown by Butler and Chen\cite{BC} and discussed in the introductory section. 
It has the structure as given
in (\ref{A_2}) and corresponds to the weak current coupling directly to the four-nucleon vertex.
In a more fundamental theory it could be due to weak interactions via virtual pions, coupling
to excited nucleons or more general two-body operators in nuclear physics language. Obviously,
this counterterm will also modify the numerical result for the fusion rate in addition to
softening the $\mu$-dependence.

\begin{figure}[ht]
   \begin{center}
          \epsfig{figure=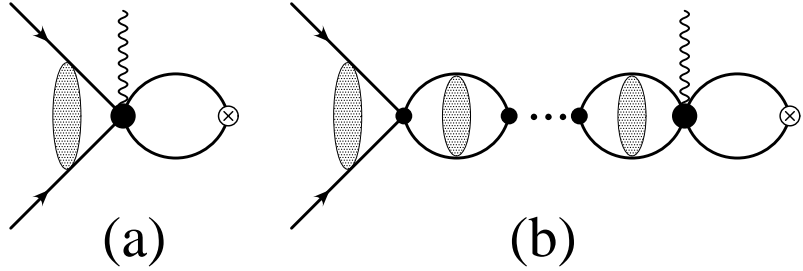,height=35mm}
   \end{center}
   \vspace{-4mm}            
   \caption{\small Contributions to the fusion amplitude from the counterterm.}
   \label{fig.9}
\end{figure}
In our case it gives a contribution depicted by the Feynman diagram in Fig.9a. It is similar
to the previously calculated contribution from Fig.8a in (\ref{Ta1}) and becomes
\beq
       T_{a}^{\,ct} = L_{1A}\int\!{d^3k\over(2\pi)^3} \int\!{d^3q\over(2\pi)^3} 
              {M\over \bk^2 + \gamma^2}\psi_\bp(\bq)  = - L_{1A}C_\eta I_0(\gamma)     \label{Count1}
\eeq    
The strong interactions in the initial state, now to lowest order in the derivative expansion, gives
the series of diagrams shown in Fig.9b. They form again an infinite geometric series whose sum 
\beq
       T_{b}^{\,ct} =  - L_{1A}\,C_\eta I_0(\gamma){C_0^pJ_0(p)\over 1 - C_0^pJ_0(p)}      \label{Count2}
\eeq
is given by the proton-proton physical scattering length $a_p$ from (\ref{C0p}) and 
$a(\mu)$ from (\ref{C0pmu}) in the fusion limit $p \ra 0$. With the regularized value for the integral $I_0$,
we thus find the total contribution from the counterterms to be
\beq
       T^{\,ct} =  - L_{1A}\,C_\eta {M\over 4\pi}(\mu - \gamma)\left[\mu - {1\over a(\mu)}\right]a_p
\eeq
The corresponding reduced matrix element then follows from (\ref{Lambda}) after multiplication by the 
wavefunction renormalization constant $Z_0$. 

We now include this new contribution as a correction to the matrix element (\ref{L2}) coming from the
ordinary axial current interactions. For the combined result we then have
\beq
    \Lambda_2^{ct}(0) &=& \Lambda_0(0)\left[1 + {1\over 2}\gamma\rho_d\right]
                  -  {a_p\gamma^2\over 4\pi}(\mu - \gamma) \left[\mu - {1\over a(\mu)}\right]\nn \\
&\times&\left[L_{1A}(\mu) - {M\over 2}\left(C_2^p(\mu) + C_2^d(\mu)\right)\right]   \label{L2fin}
\eeq
The coupling constant $L_{1A}$ of the counterterm must have a dependence on the renormalization scale
$\mu$ so that the total $\mu$-dependence in the last term is negligible. When $\mu \gg \gamma$ we see
that this requirement leads to
\beq
     L_{1A}(\mu)_{\mu \gg \gamma} = {4\pi \ell_{1A}\over M\mu^2}                    \label{L1A}
\eeq
where $\ell_{1A}$ is an unknown, dimensionless constant. It is set by physics on scales shorter than
$1/m_\pi$ and its natural value should be around one as pointed out by Butler and Chen\cite{BC}. 
In order to get a rough idea of the sensitivity
of the result on this parameter, we take $\mu = m_\pi$ which is the scale at which one should match
the effective theory to the more fundamental theory involving pions. Varying then $\ell_{1A}$ in the
interval $[-1,1]$, we find that the fusion rate measured by $\Lambda_2^{ct}(0)^2$ varies linearly from
6.22 to 6.84. These values are seen to be systematically below the effective-range result following 
from (\ref{ER}), but are within the 5\% - 10\% uncertainty range expected at this order.

\section{Discussion and conclusion}
Effective field theory is a very powerful approach to low-energy phsyics. It can hardly be said to be wrong
when used correctly since it is just based upon the basic symmetries of the problem and standard quantum
field theory. In that way it is a very conservative approach since it does not admit assumptions about
the physics on scales shorter than it is meant to handle. Instead of such specific and model-dependent
assumptions, one has higher-dimensional contact interactions and counterterms 
with coupling constants which represent the unknown physics. The most common criticism against effective 
field theory is therefore that it is not accurate enough since the results may depend on one or more 
such coupling constants which are not {\it a priori} known. One
can make estimates of these unknown coupling constants based upon some kind of naturalness supported by
dimensional analysis and the renormalization group.

But these counterterms do not really represent a weakness of effective field theory. Since they are
interactions appearing in a Lagrangian, they will appear with the same strength in many different
processes. If one or more of these allow for the determination of the corresponding coupling
constants, one can then make much more accurate predictions for the other reactions. One recent example is
radiative neutron-proton  capture $n + p \ra d + \gamma$. When the process takes place at very low 
energies or at rest, it is dominated by a magnetic dipole transition which at next-to-leading order
also involves a four-nucleon counterterm very similar to the one we have considered here for proton-proton
fusion. From the measured rate at these low energies, the counterterm can then be determined
numerically\cite{CRS}. The same neutron-proton fusion process is also a key reaction
in big-bang nucleosynthesis where it takes place at energies upto around 1 MeV. Chen and Savage have
now calculated the corresponding cross section with an uncertainty of 4\% based on the measured
counterterm\cite{CS}. A similar accuracy can be expected also for proton-proton fusion if the counterterm
can be determined in some other process.

It has already been pointed out that our results for the proton-proton fusion  have a
very similar structure to what one finds in the effective-range approximation in nuclear physics.
This has also been seen in other processes investigated within the 
same effective theory and at higher orders in the perturbative expansion\cite{KSW_2}\cite{CRS}. It is
understood when one realizes that these processes are dominated by the properties of the deuteron 
wavefunction at large distance scales which is contained in the effective-range approximation.
In the KSW field theory, these properties are coded into the coupling constants $C_0^d$ and $C_2^d$. 
While $C_0^d$ is responsible for
binding the deuteron and must be treated non-perturbatively, the effects of $C_2^d$ are to be treated
perturbatively and gives the detailed behaviour of the wavefunction at large distances. In the above
$C_2^d$ was determined by matching to the effective range parameter $\rho_d$. In order to get better
agreement with low-energy proton-neutron scattering data which are related directly to the deuteron 
bound state wavefunction via analytical continuation, it has recently been pointed out by Phillips, Rupak
and Savage that one should instead match $C_2^d$ to the wavefunction normalization parameter 
$Z_d$\cite{PRS}. This gives the result
\beq
     C_2^d(\mu) = {2\pi\over\gamma M} {Z_d - 1\over (\mu - \gamma)^2}
\eeq  
where $Z_d = 1.69$. They have shown that this markedly improves the convergence of the perturbative
calculation of many processes involving deuterons at low energies. Rupak has recently applied this 
improved method to  neutron-proton fusion $n + p \ra d + \gamma$ at
energies relevant to big-bang nucleosynthesis as discussed above\cite{Gautam}. Including one higher order 
in the perturbative expansion of the elctric transition amplitude, he has then obtained an accuracy of 
1\% for the calculated cross section.

In our case we can now use this
new value for $C_2^d$ in the result (\ref{L2}) for the reduced matrix element. Including also the 
counterterm as in (\ref{L2fin}), we then obtain our final result. Again  the counterterm coupling constant
will have the form (\ref{L1A}) for large values of the renormalization
mass. Choosing $ \mu = m_\pi$, we now find that $\Lambda_2^2(0)$ varies between 7.04 and 7.70 when
the parameter  $\ell_{1A}$ takes values in the interval $[-1,1]$. With the size of the unknown counterterm
in this range, we thus have the central value $\Lambda_2^2(0) = 7.37$ with a
conservative estimate for the uncertainty of 6\% - 8\%.
We thus find a somewhat higher value for the fusion rate in this improved perturbative calculation 
compared with results from effective range theory (\ref{ER}) and the inclusion of axial two-body 
effects\cite{Stoks}. But these other results are now within the accuracy range of the effective field 
theory method. Should there in the future turn out
to be a real discrepancy between these different theoretical descriptions, it can only be due to meson 
processes at shorter scales which have been overlooked or not correctly handled in the more conventional 
nuclear physics approach.

The only way to improve the accuracy and thus obtain a more predictive result, is
to establish the value of the counterterm coupling constant $L_{1A}$. In principle it could be measured
in many other reactions, but the most promising is elastic and inelastic scattering of neutrinos on
deuterons as shown by Butler and Chen\cite{BC}. High-precision experimental results for these reactions
would then result in a known value of the relevant counterterm. As shown by Rupak for radiative
neutron capture, one should then be in the position to obtain the proton-proton fusion rate with a much
improved accuracy. This will place our understanding of this fundamental process on a more solid
basis. Needless to say, it will also strengthen our knowledge of the neutrino production rate in the 
Sun.

\section{Acknowledgement}
We want to thank John Bahcall, Jiunn-Wei Chen,  Peter Lepage, Gautam Rupak, Martin Savage and Mark Wise 
for encouragement and many helpful discussions. Most of this work was done in the 
Department of Physics and INT at the University of Washington in Seattle and we are grateful
for generous support and hospitality.

\section{Appendix}
We will here regularize and evaluate the divergent integrals involving Coulomb wavefunctions
which are needed for the effective-range corrections to the fusion rate. Some of them have previously 
been encountered in connection with higher order corrections
to low-energy proton-proton elastic scattering\cite{KR_pp}. They were then calculated by a method based
on regularization of the Fourier-transformed Coulomb wavefunctions. We will here use a different and 
simpler method. 

The simplest integral is $J_2(p)$ in (\ref{J2p}) which we rewrite as
\be
    J_2(p) = p^2 J_0(p) +  \int\!{d^3k\over(2\pi)^3} \int\!{d^3q\over(2\pi)^3}
            (\bk^2 - \bp^2)\, \bra{\bk}G_C(E)\ket{\bq}
\ee    
It represents a Coulomb-dressed bubble propagator with a derivative interaction at one vertex. 
Here we have introduced the free eigen-momentum states $\bra{\bk}$ and $\ket{\bq}$. The Coulomb
propagator $G_C(E)$ satifies the Lippmann-Schwinger equation  $G_C = G_0 + G_0V_CG_C$ where $G_0(E)$
is the free propagator (\ref{G0}) and $V_C$ is the Coulomb potential. In momentum space it has the 
matrix element $\bra{\bk}V_C\ket{\bk'} = 4\pi\alpha/(\bk - \bk')^2$. The first term will now give zero 
with the use of dimensional regularization,
\beq
     \int\!{d^dk\over(2\pi)^d} = 0
\eeq
We then insert two complete sets of momentum eigenstates between the three operators in the matrix 
elements in the second term. The denominator in the free propagator $G_0$ then cancels against
the factor $\bk^2 - \bp^2$ in the integral. We are thus left with
\be
   \int\!{d^3k\over(2\pi)^3}(\bk^2 - \bp^2)\,\bra{\bk}G_0V_CG_C\ket{\bq}
  = -M\int\!{d^3k\over(2\pi)^3}\int\!{d^3k'\over(2\pi)^3}
   {4\pi\alpha\over(\bk - \bk')^2}\bra{\bk'}G_C(E)\ket{\bq}
\ee
In the integral over $\bk$ we now shift the integration variable $\bk \ra \bk + \bk'$ and use the PDS 
regularization result
\beq
         \int\!{d^3k\over(2\pi)^3}{4\pi\alpha\over \bk^2} = \alpha\mu            \label{PDSreg}
\eeq
The remaining two integrals over $\bk'$ and $\bq$ then simply gives $J_0(p)$. We thus have the result
\beq
     J_2(p) = [p^2 - \alpha M\mu]\,J_0(p)
\eeq
Except for a higher order term in the fine-structure constant $\alpha$, this agrees with what we obtained
with the much more cumbersome wavefunction regularization method\cite{KR_pp}.

The next integral $\psi_2(p)$ in (\ref{psi2}) corresponds to the double drivative of the Coulomb
wavefunction at the origin. We can write it as
\be
    \psi_2(p) = p^2\psi_0(p) 
              + \int\!{d^3k\over(2\pi)^3}(\bk^2 - \bp^2)\, {\mbox{$\langle\bk\,$}}\ket{\psi_\bp}
\ee    
where $\ket{\psi_\bp}$ is a Coulomb state with momentum $\bp$. It can formally be expressed in terms of 
the free state $\ket{\bp}$ as
\be
    \ket{\psi_\bp} = [1 + G_CV_C]\ket{\bp}
\ee
One then has 
\be
   \int\!{d^3k\over(2\pi)^3}(\bk^2 - \bp^2)\, {\mbox{$\langle\bk\,$}}\ket{\psi_\bp}
   &=& \int\!{d^3k\over(2\pi)^3}(\bk^2 - \bp^2)\, \left[\bra{\bk}G_0V_C\ket{\bp} +
       \bra{\bk}G_0V_CG_CV_C\ket{\bp}\right] \\
   &=& \int\!{d^3k\over(2\pi)^3}(\bk^2 - \bp^2)\,\bra{\bk}G_0V_C\ket{\psi_\bp}
\ee
using $G_CV_C\ket{\bp} = \ket{\psi_\bp} - \ket{\bp}$ in the last term. Inserting now again 
two complete sets of free momentum states as above, it follows that
\be
    \psi_2(p) = p^2\psi_0(p) - M\int\!{d^3k\over(2\pi)^3}\int\!{d^3q\over(2\pi)^3}
                {4\pi\alpha\over (\bk - \bq)^2}{\mbox{$\langle\bq\,$}}\ket{\psi_\bp}
\ee
After a shift of integration variable, we have the result
\beq
     \psi_2(p) = [p^2 - \alpha M\mu]\,\psi_0(p)
\eeq
when making use the the PDS regularized integral (\ref{PDSreg}).

The last integral we need is $B_2'(p)$ in (\ref{B2prim}). Rewriting it as above, it takes the form
\be
    B_2'(p) = p^2 B_0(p) + M \int\!{d^3k\over(2\pi)^3}{d^3q\over(2\pi)^3} 
           {\bq^2 - \bp^2\over \bk^2 + \gamma^2}G_C(E;\bk,\bq)                     
\ee
where the first term is the finite integral (\ref{B0}). In the second term we can use the 
Lippmann-Schwinger equation for the Coulomb propagator. Again we find that the denominator of the free 
propagator cancels against $\bq^2 - \bp^2$ in the numerator. The first term in the integral
gives then just the integral $I_0(p)$ in (\ref{I0}). Going through the same steps as above with
insertion of complete sets of states, the second term is then reduced to the finite integral $B_0(p)$. In
this way we obtain
\beq
    B_2'(p) = MI_0(p) + [p^2 - \alpha M\mu]\,B_0(p)                             \label{B2prime}
\eeq
which again is a surprising simple result.

We notice that these three divergent Coulomb integrals contain the common factor $p^2 - \alpha M\mu$ in
the results. This can be understood as coming from the divergence of the double derivative of the 
Coulomb wavefunction $\psi_\bp(\br)$ at the origin. It satisfies the Sch\"odinger wave equation
\be
     \left[-{1\over M}{\bg\nabla}^2 + V_C(r)\right]\psi_\bp(\br) = E\psi_\bp(\br)
\ee
where the energy $E = p^2/M$. When we now take the limit $r \ra 0$, it follows that
\beq
      -{\bg\nabla}^2 \psi_\bp(\br)_{r\ra 0} = [p^2 - \alpha M\mu]\,\psi_\bp(0)
\eeq
since the regularized integral (\ref{PDSreg}) is just the Coulomb potential at the origin.

\end{document}